\documentclass[preprint,aps,showpacs]{revtex4}

\usepackage{graphicx}
\usepackage{bm}       
\usepackage{mathptmx} 
\usepackage{amsmath}
\usepackage{amssymb}
\begin{document}

\def\slashchar#1{\setbox0=\hbox{$#1$}           
   \dimen0=\wd0                                 
   \setbox1=\hbox{/} \dimen1=\wd1               
   \ifdim\dimen0>\dimen1                        
      \rlap{\hbox to \dimen0{\hfil/\hfil}}      
      #1                                        
   \else                                        
      \rlap{\hbox to \dimen1{\hfil$#1$\hfil}}   
      /                                         
   \fi}                                         %
 

\title{Quark mass dependence of two-flavor QCD}
\author{Michael Creutz}
\affiliation{Physics Department, Brookhaven National Laboratory\\
Upton, NY 11973, USA
}
                      
\begin{abstract}{ I explore the rich phase diagram of two-flavor QCD as a
  function of the quark masses.  The theory involves three parameters,
  including one that is CP violating.  As the masses vary, regions of
  both first and second order transitions are expected.  For
  non-degenerate quarks, non-perturbative effects cease to be
  universal, leaving individual quark mass ratios with a
  renormalization scheme dependence.  This raises complications in
  matching lattice results with perturbative schemes and demonstrates
  the tautology of attacking the strong CP problem via a vanishing up
  quark mass.}
\end{abstract}


\maketitle

\section{Introduction}
The standard theory of the strong interactions is based on quarks
interacting through non-Abelian gauge fields.  This system is
remarkable in its paucity of parameters.  Once the overall scale is
set, perhaps by working in units where the proton mass is unity, the
only remaining parameters are the quark masses.  In general these are
complex numbers, although field redefinitions allow removing all
phases but one, usually called the strong CP parameter Theta. For a
recent review, see Ref.~\cite{Vicari:2008jw}. Thus the number of
physical parameters for QCD is one more than the number of quark
species.  As is well known, if Theta is non-trivial, the theory
violates CP symmetry.  As CP appears to be a good symmetry of hadronic
physics, the strong CP puzzle asks the question why should this
parameter be so small experimentally.

In this paper I restrict myself to two-flavor QCD and explore the
qualitative behavior as the most general mass terms are varied.  Using
effective potential techniques, I find a rich phase diagram with
regions of both first and second order phase transitions.  I find that
there can be interesting long distance physics even when no individual
quark mass vanishes.  I also delve more deeply into the old argument
\cite{Creutz:2003xc} for a fundamental ambiguity in defining a
vanishing quark mass.  These effects are inherently non-perturbative
and lead to unsettled issues for matching lattice with perturbative
results.

Of course, with QCD being an interacting quantum field theory, nothing
has been proven rigorously.  To proceed I assume that QCD exists as a
field theory and confines in the usual way.  In addition I will work
in the conventional picture of spontaneous breaking of approximate
chiral symmetry as the explanation for the lightness of the pions.  I
also assume the generation of the singlet pseudoscalar meson mass is
tied to the anomaly.  For simplicity I work with the two-flavor theory
with only the $u$ and $d$ quarks, assuming their masses are light
enough that conventional chiral expansions make sense.  The
generalization to more flavors is straightforward, although there are
some rather fascinating further consequences \cite{Creutz:2003xu}.

I begin in Section \ref{spinflip} with a simple argument on how the
various quark masses indirectly influence each other.  The obscurity
of these effects in a mass independent regularization scheme has
raised some controversy, which I address in Section \ref{critique}.
Section \ref{general} turns to the most general mass term for the
two-flavor theory.  Here I discuss some of the conventions needed for
formulating this question.  Section \ref{strongcp} relates the mass
parameters to the strong CP problem and discusses the issues with
pursuing a vanishing lightest quark mass.  Section \ref{diagram} uses
an effective potential argument to develop the qualitative phase
diagram as a function of the independent mass parameters. Finally, the
basic ideas are summarized in section \ref{summary}.

\section{Spin-flip quark scattering}
\label{spinflip}

I begin with a reminder of some basic properties expected for massless
two-flavor QCD.  While the classical theory is conformally invariant,
it is commonly believed that in the quantized theory confinement and
dimensional transmutation generate a non-trivial mass scale
$\Lambda_{qcd}$.  This scale is scheme dependent, but that will not
enter the qualitative discussion here.  In particular, the theory
should contain massive stable nucleons.  On the other hand,
spontaneous chiral-symmetry breaking is expected to give rise to three
massless pions as Goldstone bosons.  In addition, the two-flavor
analog of the eta prime meson should acquire a mass from the anomaly.

In this picture, the eta prime and neutral pion involve distinct
combinations of quark-antiquark bound states.  In the simple quark
model the neutral pseudoscalars involve the combinations
\begin{align}
&\pi_0 \sim \overline u \gamma_5 u -\overline d\gamma_5 d\\
&\eta^\prime \sim \overline u\gamma_5 u
 +\overline d\gamma_5 d +\hbox{glue}.
\end{align}
Here I include a gluonic contribution from mixing between the
$\eta^\prime$ and glueball states.  When the quarks are degenerate,
isospin forbids such mixing for the pion.

\begin{figure}
\centering
{\includegraphics[width=3in]{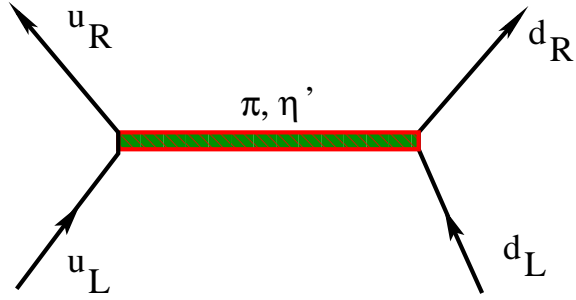}}
\caption{Both pion and eta prime exchange can contribute to spin flip
  scattering between up and down quarks. }
\label{scattering}
\end{figure}

Projecting out helicity states for the quarks,
$q_{R,L}=(1\pm\gamma_5)q/2$, the pseudoscalars are combinations of
left with right states, {\it i.e.} $\overline q_L q_R-\overline q_R
q_L$.  Thus, as shown schematically in Fig.~\ref{scattering}, meson
exchange will contribute to a hypothetical quark-quark spin-flip
scattering experiment.  More precisely, the four point function
$\langle \overline u_R u_L \overline d_R d_L\rangle$ should not
vanish.  (Scalar meson exchange will also contribute to this process,
but this is not important for the qualitative argument below.)  Of
course I assume that some sort of gauge fixing has been done to
eliminate a trivial vanishing of this function from an integral over
gauges.

It is important that the $\pi_0$ and $\eta^\prime$ are not degenerate.
This is due to the anomaly and the fact that the $\eta^\prime$ is not
a Goldstone boson.  At a more abstract level this
$\pi_0$--$\eta^\prime$ splitting is ascribed to topological structures
in the gauge field, but such details are not necessary for the
discussion here.  Because the mesons are not degenerate, their
contributions to the above diagram cannot cancel.  The conclusion of
this simple argument is that helicity-flip quark-quark scattering is
not suppressed as the mass goes to zero.

\begin{figure}
\centering
\includegraphics[width=3in]{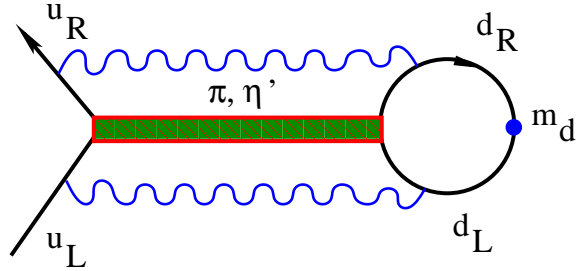}
\caption{Through physical meson exchange, a down quark mass can induce
  an effective mass for the up quark.  The gluon exchanges can
  compensate for the pseudoscalar nature of the meson fields.}
\label{induced}
\end{figure}

Now consider turning on a small {$d$} quark mass while leaving the up
quark massless.  Formally this mass allows one to connect the in-going
and out-going down-quark lines in Fig.~\ref{scattering} and thereby
induce a mixing between the left and right handed up quark.  Such a
process is sketched in Fig.~\ref{induced}.  Here I allow for
additional gluon exchanges to compensate for turning the pseudoscalar
field into a traditional mass term.

So the presence of a non-zero {$d$} quark mass will induce an
effective mass for the {$u$} quark, even if the latter initially
vanishes. As a consequence, non-perturbative effects will renormalize
{$m_u/ m_d$}.  If this ratio is zero at some scale, it cannot remain
so for all scales.  Only in the isospin limit are quark mass ratios
renormalization group invariant.  As lattice simulations include all
perturbative and non-perturbative effects, this phenomenon is
automatically included in such an approach.

Confinement plays a crucial role in what is effectively an ambiguity
in defining quark masses.  Because quarks cannot travel long distances
in isolation, their masses cannot be directly inferred from long
distance propagators.  This is tied directly with the phase diagram
discussed in Section \ref{phasediagram}, where it is shown that no
discernable physical structure is seen when single quark mass
vanishes.

This cross talk between the masses of different quark species is a
relatively straightforward consequence of the chiral anomaly and has
been discussed several times in the past, usually in the context of
gauge field topology and the index theorem \cite{
  Georgi:1981be,Banks:1994yg, Creutz:2003cj, Creutz:2003xc}.  Despite
the simplicity of the above argument, the conclusion is frequently met
with skepticism from the perturbative community.  In perturbation
theory, spin flip processes are suppressed as the quark masses go to
zero.  The above discussion shows that this lore need not apply when
anomalous processes come into play.  In particular, mass
renormalization can not be flavor blind and the concept of mass
independent regularization is problematic.  Since the quark masses
influence each other, there are inherent ambiguities defining $m_u=0$.
This has consequences for the strong CP problem, discussed further
below.  Furthermore, since these effects involve quark mass
differences, a traditional perturbative regulator such as
$\overline{MS}$ is not complete when $m_u\ne m_d$.  Because of this,
the practice of matching lattice calculations to $\overline{MS}$ is
problematic, a point that is sometimes ignored
\cite{Davies:2009ih,Blum:2010ym}.  (Ref. \cite{Davies:2009ih} also
suffers from an uncontrolled extrapolation in the number of quark
species \cite{Creutz:2009zq}.)

\section{Specific critiques}
\label{critique}

Given the simplicity of the argument in the previous section, it may
seem surprising that it often receives severe criticism.  The first
complaint sometimes made is that one should work directly with bare
quark masses.  This ignores the fact that the bare quark masses all
vanish under renormalization.  The renormalization group equation for
a quark mass reads
\begin{equation}
a {dm_i\over da}=\gamma(g) m_i = \gamma_0 g^2 + O(g^4)
\label{mrg}
\end{equation}
where the leading coefficient is well known, $\gamma_0={8\over (4\pi)^2}.$
As asymptotic freedom drives the bare coupling to zero, the bare
masses behave as
\begin{equation}
m\sim g^{\gamma_0/\beta_0} (1+O(g^2))\rightarrow 0
\label{mflow}
\end{equation}
where $\beta_0$ (explicitly given later) is the first term in the beta
function controlling the vanishing of the bare coupling in the
continuum limit.  Since all bare quark masses are formally zero, one
must address these questions in terms of a renormalization scheme at a
finite cutoff.

The second objection often made is that in a mass independent
regularization scheme, mass ratios are automatically constant.  Such
an approach asks that the renormalization group function $\gamma(g)$
in Eq.~(\ref{mrg}) be chosen to be independent of the quark species
and mass.  This immediately implies the constancy of all quark mass
ratios.  As only the first term in the perturbative expansion of
$\gamma(g)$ is universal, a mass independent scheme is indeed an
allowed procedure.  However, such a scheme obscures the off-diagonal
$m_d$ effect on $m_u$ discussed above.  In particular, by forcing
constancy of bare mass ratios, one will find that the ratios of
physical particle masses will vary as a function of cutoff.  This will
be in a manner that cancels the flow from the process in Section
\ref{spinflip}.  The fact that physical particle mass ratios are not
just a function of quark mass ratios is shown explicitly in Section
\ref{phasediagram}, where it is shown that in the chiral limit the
combination $1-{m_{\pi_0}^2 / m_{\pi_\pm}^2}$ is proportional to
$(m_d-m_u)^2\over (m_d+m_u)\Lambda_{qcd}$.

From a non-perturbative point of view, having physical mass ratios
vary with cutoff seems rather peculiar; indeed, the particle masses
are physical quantities that would be natural to hold fixed.  And,
even though a mass independent approach is theoretically possible,
there is no guarantee that any given ratio ${m_i\over m_j}$ will be
universal between schemes.  Finally, the lattice approach itself is
usually implemented with physical particle masses as input.  As such
it is not a mass independent regulator, making a perturbative matching
to lattice results rather subtle.

A third frequent complaint against the argument in Section
\ref{spinflip} is that one should simply do the matching at some high
energy, say 100 GeV, where ``instanton'' effects are exponentially
suppressed and irrelevant.  This point of view has several problems.
First, the lattice simulations are not done at miniscule scales and
non-perturbative effects are present and substantial.  Furthermore,
the exponential suppression of topological effects is in the inverse
coupling, which runs logarithmically with the scale.  As such, the
non-perturbative suppression is a power law in the scale and
straightforward to estimate.

Recall the renormalization group prediction for how the
eta prime mass depends on the coupling in the continuum limit
\begin{equation}
\label{rg}
m_{\eta^\prime} \propto {1\over a}{ e^{-1/(2\beta_0 g^2)}
g^{-\beta_1/\beta_0^2}}.
\end{equation}
Here $\beta_0= {11-2n_f/3 \over (4\pi)^2}$, $\beta_1={102-12n_f\over
  (4 \pi)^4}$, $n_f$ is the number of quark flavors, and $a$ is the
cutoff scale, {\it i.e.} the lattice spacing with such a
renormalization scheme.  While this formula indeed shows the
exponential suppression in $1/g^2$, this is cancelled by the inverse
cutoff factor in just such a way that the mass of this physical
particle remains finite.  The ambiguity in the quark mass splitting is
controlled by the mass splitting $m_{\eta^\prime}-m_{\pi_0}$ as well
as being proportional to $m_d-m_u$.  Considering $m_d=5$ MeV at a
scale of $\mu=2$ GeV, a rough estimate of the order of the $u$ quark
mass shift is
\begin{equation}
\Delta m_u(\mu) \sim 
\left({m_{\eta^\prime}-m_{\pi_0}\over \Lambda_{qcd}}\right)
\ (m_d-m_u)
= O(1\ \hbox{MeV}),
\end{equation} 
a number comparable to typical phenomenological estimates.  Of course
the result depends on scale, but that dependence is only logarithmic
and given by Eq.~(\ref{mflow}).  Additional flavors will reduce the
size of this effect; with the strange quark present, it should be
proportional to $m_d m_s$.

It is important to note that for a modest number of flavors the
exponent controlling the coupling constant suppression in Eq.~\ref{rg}
differs substantially from the classical instanton action
\begin{equation}
{ 8\pi^2\over (11-2n_f/3) g^2}<< 
{ 8\pi^2\over g^2}.
\end{equation} 
This difference arises because one should consider topological
excitations above the quantum, not the classical, vacuum.  Zero modes
of the Dirac operator are still responsible for the bulk of the eta
prime mass, but naive semi-classical arguments strongly underestimate
their effect.

\section{General masses in two-flavor QCD}
\label{general}

Given the confusion over the meaning of quark masses, it is useful to
explore the behavior of two-flavor QCD as these quantities are varied.
Here I review how the theory depends on the three non-trivial mass
parameters.  These includes the possibility of explicit CP violation.
The full theory has a rather rich phase diagram, including both first
and second order phase transitions, some occurring when none of the
quark masses vanish.

For the following the quark fields $\psi$ carry implicit isospin,
color, and flavor indices.  I assume that the theory in the massless
limit has the flavored chiral symmetry under
\begin{eqnarray}
\psi \longrightarrow e^{i\gamma_5 \tau_\alpha\phi_\alpha/2}\psi\cr
\overline \psi \longrightarrow \overline\psi 
e^{i\gamma_5 \tau_\alpha\phi_\alpha/2}.
\end{eqnarray}
Here $\tau_\alpha$ represents the Pauli matrices generating isospin
rotations.  The angles $\phi_\alpha$ are arbitrary rotation
parameters.  This, of course, is the chiral symmetry that is
spontaneously broken to give the massless Goldstone pions.

I wish to construct the most general possible two-flavor mass term to
add to the massless Lagrangian.  Such should be a dimension 3
quadratic form in the fermion fields and should transform as a singlet
under Lorentz transformations.  For simplicity, I only consider
quantities that are charge neutral as well.  This leaves four
candidate fields, giving the generalized form for consideration
\begin{equation}
m_1\overline\psi\psi+
m_2\overline\psi\tau_3\psi+
im_3\overline\psi\gamma_5\psi+
im_4\overline\psi\gamma_5\tau_3\psi.
\label{genmass}
\end{equation}
The first two terms are naturally interpreted as giving the average
quark mass and the quark mass difference, respectively.  The remaining
two terms are less conventional.  The $m_3$ term is connected with the
CP violating parameter of the theory.  The final $m_4$ term has been
used in conjunction with the Wilson discretization of lattice
fermions, where it is referred to as a ``twisted mass''
\cite{Frezzotti:2000nk,Boucaud:2007uk}. Its utility in that context is
the ability to reduce lattice discretization errors, but that is not
the subject of this note.

These four terms are not independent.  Indeed, consider the above
flavored chiral rotation in the $\tau_3$ direction, $\psi\rightarrow
e^{i\theta\tau_3\gamma_5}\psi$.  Under this the terms transform as
\begin{eqnarray}
\overline\psi\psi\ &\longrightarrow\  
\cos(\theta)\overline\psi\psi
+\sin(\theta)i\overline\psi\gamma_5\tau_3\psi\cr
\overline\psi\tau_3\psi\ &\longrightarrow\
\cos(\theta)\overline\psi\tau_3\psi
+\sin(\theta)i\overline\psi\gamma_5\psi\cr
i\overline\psi\tau_3\gamma_5\psi\ &\longrightarrow\  
\cos(\theta)i\overline\psi\tau_3\gamma_5\psi
-\sin(\theta)\overline\psi\psi\cr
i\overline\psi\gamma_5\psi\ &\longrightarrow\
\cos(\theta)i\overline\psi\gamma_5\psi
-\sin(\theta)\overline\psi\tau_3\psi
\label{flavorrot}
\end{eqnarray}
Such a rotation mixes $m_1$ with $m_4$ and $m_2$ with $m_3$.
Using this freedom, one can select any one of the $m_i$ to vanish and a
second to be positive.

The most common choice is to set $m_4=0$ and use $m_1$ as controlling
the average quark mass.  Then $m_2$ gives the quark mass difference,
and CP violation appears in $m_3$.  This, however, is only a
convention.  The alternative ``twisted mass'' scheme
\cite{Frezzotti:2000nk,Boucaud:2007uk}, makes the choice $m_1=0$.
This uses {$m_4>0$} for the average quark mass, and {$m_3$} becomes the
up-down mass difference.  In this case $m_2$ becomes the CP violating
term.  It is amusing to note that an up down quark mass difference in
this formulation involves the naively CP odd
$i\overline\psi\gamma_5\psi$.  The strong CP problem has been rotated
into the smallness of the $\overline\psi\tau_3\psi$ term, which with
the usual conventions is the mass difference.  But because of the
flavored chiral symmetry, both sets of conventions are physically
equivalent.

For the following I make the arbitrary choice $m_4=0$, although one
should remember that this is only a convention and I could have chosen
any of the four parameters in Eq.~(\ref{genmass}) to vanish.  With
this choice two-flavor QCD, after scale setting, depends on three mass
parameters
\begin{equation}
m_1\overline\psi\psi+
m_2\overline\psi\tau_3\psi+
im_3\overline\psi\gamma_5\psi.
\end{equation}
It is the possible presence of $m_3$ that represents the strong CP
problem.  As all the parameters are independent and transform
differently under the symmetries of the problem, there is no
connection between the strong CP problem and $m_1$ or $m_2$.

As is well known, the chiral anomaly is responsible for the singlet
rotation
\begin{eqnarray}
\psi \longrightarrow e^{i\gamma_5 \phi/2}\psi\cr
\overline \psi \longrightarrow \overline\psi 
e^{i\gamma_5 \phi/2}
\end{eqnarray}
not being a valid symmetry, despite the fact that $\gamma_5$ naively
anti-commutes with the massless Dirac operator.  The anomaly is quite
nicely summarized via Fujikawa's \cite{Fujikawa:1979ay} approach where
after the above rotation the fermion measure in the path integral
picks up a factor of
\begin{equation}
\det( e^{i\gamma_5 \phi})=\exp(i\phi{\rm Tr}\ \gamma_5).
\end{equation}
Using the Dirac operator $\slashchar D$ itself as a regulator, define
\begin{equation}
{\rm Tr}\ \gamma_5=\lim_{\Lambda\rightarrow\infty}\gamma_5
e^{\slashchar D^2/\Lambda^2}.
\end{equation}
In any give gauge configuration only the zero eigenmodes of
$\slashchar D$ contribute, and by the index theorem this is connected
to the winding number of the gauge configuration.  The conclusion is
that the above rotation changes the fermion measure by an amount
depending non-trivially on the gauge field configuration.

Note that this anomalous rotation allows one to remove any topological
term from the gauge part of the action.  Naively this would have been
yet another parameter for the theory, but by including all three mass
terms for the fermions, this can be absorbed.  For the following I
consider that any topological term has thus been rotated away. After
this one is left with the three mass parameters above, all of which
are independent and relevant to physics. 

These parameters are a complete set for two-flavor QCD; however, this
choice differs somewhat from what is often discussed.  Formally one
defines the more conventional variables as
\begin{align}
 &m_u=m_1+m_2+im_3\cr
 &m_d=m_1-m_2+im_3\cr
&e^{i\Theta}={m_1^2-m_2^2-m_3^2+2im_1m_3\over
\sqrt{m_1^4+m_2^4+m_3^4+2m_1^2m_3^2+2m_2^2m_3^2-2m_1^2m_2^2
}}.
\end{align}
Particularly for $\Theta$, this is a rather complicated change of
variables.  For non-degenerate quarks in the context of the phase
diagram discussed below, the variables $\{m_1,m_2,m_3\}$ are more
natural.

\section {The strong CP problem}
\label{strongcp}
The strong interactions preserve CP to high accuracy.  Thus only two
of the three possible mass parameters seem to be needed. With the
above conventions, it is natural to ask why is $m_3$ so small?

It is the concept of unification that brings this question to the
fore.  The weak interactions of course do violate CP.  Thus, if the
electroweak and the strong interactions separate at some high scale,
why doesn't some remnant of this breaking survive in the strong
sector?  How is CP recovered for the nuclear force?

Several ``solutions'' to this puzzle have been proposed.  Perhaps the
simplest is that there is no unification and the strong interactions
should be considered on their own with the electroweak effects being
only a small perturbation.  A second approach is to add an additional
``axion'' field to make the CP phase a dynamical field that relaxes to
zero \cite{Peccei:1977hh,Weinberg:1977ma}.  The coupling of this
additional field is not determined a priori, and thus it need only be
small enough to have avoided detection in past experiments.

Another often proposed solution involves having the lightest quark
mass vanish, making its phase irrelevant.  Several years ago this was
criticized because the definition of an isolated quark mass is
inherently ambiguous due to confinement \cite{Creutz:2003xc}.  As this
conclusion remains controversial, I return to this topic and reexpress
the problem in terms of the above mass terms.  I hope this language
will clarify why relating a vanishing up quark mass to the strong CP
problem is a tautology.

Why is a vanishing up quark mass not a sensible approach?  From the
above, one can define the up quark mass as a complex number
\begin{equation}
m_u\equiv m_1+m_2+im_3
\end{equation}
But the quantities $m_1$, $m_2$, and $m_3$ are independent parameters
with different symmetry properties.  With our conventions, $m_1$
represents an isosinglet mass contribution, $m_2$ is isovector in
nature, and $m_3$ is CP violating.  And, as extensively discussed
earlier, the combination $m_1+m_2=0$ is scale and scheme dependent.
The strong CP problem only requires small $m_3$. So while it may be
true formally that
\begin{equation}
 m_1+m_2+im_3=0\ \Rightarrow\ m_3=0,
\end{equation}
this would depend on scale and one might well regard this as ``not
even wrong.''

\section {Phase diagram for general quark masses}
\label{diagram}
As a function of the three mass parameters, QCD has a rather intricate
phase diagram. From simple chiral Lagrangian arguments this diagram
can be qualitatively mapped out. Ref.~\cite{Creutz:1995wf} studied
this system in the $m_2=0$ case; a first order transition is expected
along the $m_3$ axes at $m_1=0$.  In conventional notation, this
corresponds to the strong CP parameter $\Theta$ taking the value
$\pi$.  That paper, however, incorrectly speculated on the structure
for non-degenerate quarks.  In Ref. \cite{Creutz:2009kx} the picture
was generalized to several degenerate flavors and the first order
transition at $\Theta=\pi$ was shown to be generic for all $n_f>1$.
Ref.\ \cite{Creutz:2005gb} studied the phase diagram for $m_3=0$ and
showed how the isospin breaking $m_2$ term splits the chiral
transition into two second-order transitions separated by a phase with
spontaneous CP violation. These second order transitions occur where
none of the quarks are massless.

\begin{figure}
\centering
{\includegraphics[width=3.5in] {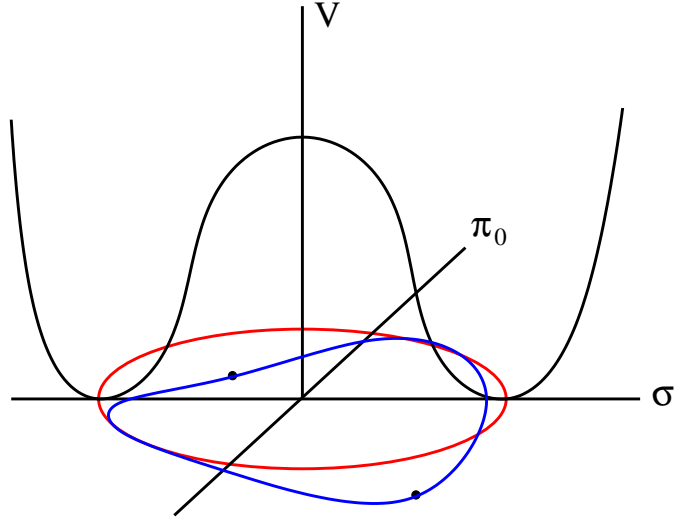}}
\caption{ The $m_2$ and $m_3$ terms warp the Mexican hat potential
  into two separate minima.  The direction of the warping is
  determined by the relative size of these parameters.}
\label{warping}
\end{figure}

The full phase diagram in terms of all mass parameters can be deduced
from a linear sigma model \cite{GellMann:1960np} analysis,
generalizing Ref.~\cite{Creutz:1995wf}.  For this, define the
composite fields
\begin{align}
&\sigma=\overline\psi\psi
&\eta^\prime=i\overline\psi\gamma_5\psi\cr
&\vec\pi=i\overline\psi\gamma_5\vec\tau\psi\qquad
&{\vec a_0}=\overline\psi\vec\tau\psi.
\end{align}
In terms of these, a natural starting effective potential is
\begin{align}
\label{potential}
V=&\lambda(\sigma^2+\vec\pi^2-v^2)^2-{m_1}\sigma-{m_2}{a_0}_3
-{m_3}\eta^\prime\cr
&+\alpha ({\eta^\prime}^2+{\vec a_0}^2)
-\beta (\eta^\prime\sigma+ {\vec a_0} \cdot \vec \pi)^2.
\end{align}
Here $\alpha$ and $\beta$ are ``low energy constants'' that bring in a
chirally symmetric coupling of $(\sigma,\vec\pi)$ with $(\eta^\prime,\vec
a_0)$.  As discussed in Ref.~\cite{Creutz:1995wf}, $\alpha$ gives mass
to the $\eta^\prime$ and $\vec a$ mesons while $\beta$ splits their masses.
The sign of the $\beta$ term is suggested so that $m_{\eta^\prime} <
m_{a_0}$.  The effect of the anomaly is manifest in these terms.

The potential in Eq.~(\ref{potential}) is a somewhat arbitrary model.
It is natural to ask if the results of this section are robust under
variations of this form.  The crucial feature of the potential is the
non-trivial minima associated with chiral symmetry breaking.
Something similar to the $\alpha$ term is needed to give the
$\eta^\prime$ a non-vanishing mass.  The $\beta$ term is somewhat
arbitrary; Ref.~\cite{Creutz:1995wf} discusses how things would change
qualitatively if it sign was reversed.  The other implicit assumption
is that the masses are small enough that they don't dramatically alter
the underlying structure of the potential.  With these caveats, the
final phase diagram should be qualitatively correct for any similar
potential.

This potential builds on the famous ``Mexican hat'' or ``wine bottle''
potential, in which the Goldstone pions are associated with the flat
directions running around at constant $\sigma^2+\vec\pi^2=v^2$.  The
$m_2$ and $m_3$ terms do not directly affect the $\sigma$ and $\pi$
fields, but induce an expectation value for ${a_0}_3$ and $\eta^\prime$,
respectively.  This in turn results in the $\alpha$ and $\beta$ terms
inducing a warping of the Mexican hat into two separate minima, as
sketched in Fig.~\ref{warping}.  The direction of this warping is
determined by the relative size of $m_2$ and $m_3$; $m_2$ ($m_3$)
warps downward in $\pi_0$ ($\sigma$) direction.  Turning on $m_1$,
this selects one of the two minima as favored.  Which one depends on
the sign of $m_1$.  This selection gives rise to a generic first order
transition at $m_1=0$.

\begin{figure}
\centering
\includegraphics[width=3.5in]{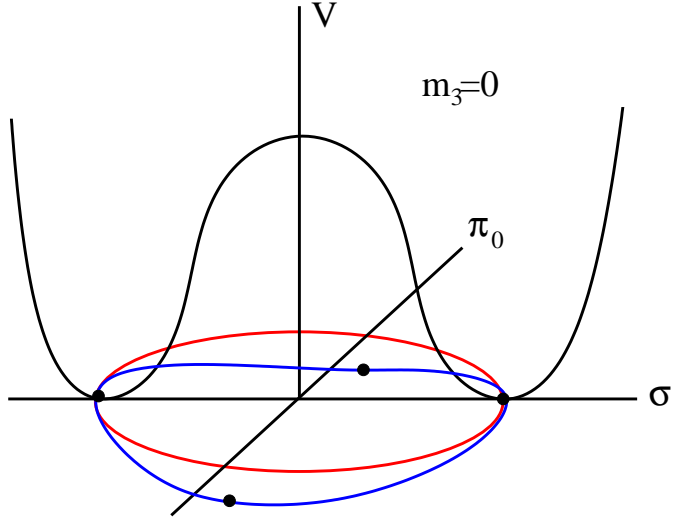} 
\caption{When $m_3=0$, the warping of the effective potential is
  downward in the $\pi_0$ direction.  The sign of $m_1$ can not pick
  one of the minima uniquely, giving the possibility of the $\pi_0$
  field spontaneously acquiring an expectation value.
}
\label{edge}
\end{figure}

\begin{figure}
\centering
\includegraphics[width=3.5in]{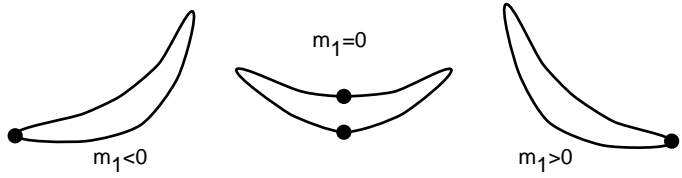} 
\caption{In the $m_1,m_2$ plane, $m_{\pi_0}^2$ can pass through zero,
giving rise to pion condensation at an Ising-like transition.  Figure
taken from \cite{Creutz:1995wf}.}
\label{ising}
\end{figure}

In addition to this transition, there is an interesting structure in
the $m_1,m_2$ plane when $m_3$ vanishes.  In this situation the
quadratic warping is downward in the $\pi_0$ direction, as sketched in
Fig.~\ref{edge}.  For large $|m_1|$ only $\sigma$ will have an
expectation, with sign determined by the sign of $m_1$.  The pion will
be massive, but the quark mass difference will give a neutral pion
mass below that of the charged pions.  As $m_1$ decreases in
magnitude at fixed $m_2$, eventually the neutral pion becomes massless
and condenses.  This is sketched in Fig.~\ref{ising}.  An order
parameter for the transition is the expectation value of the $\pi_0$
field, with the transition being in the class of the four dimensional
Ising model.

In this simple model the ratio of the neutral to charged pion masses
can be estimated from a quadratic expansion about the minimum of the
potential.  For $m_3=0$ and $m_1$ above the transition line, this
gives
\begin{equation}
{m_{\pi_0}^2\over m_{\pi_\pm}^2}
=1-{\beta v m_2^2\over 2\alpha^2 m_1}+O(m^2).
\end{equation}
The second order transition is located where this vanishes, and thus
occurs for $m_1$ proportional to $m_2^2$.  Note that this equation
shows that a constant quark mass ratio does not correspond to a
constant meson mass ratio and vice versa.  This is the ambiguity
discussed in Section \ref{spinflip}.  This model should not be trusted
when the quark masses become of order $\Lambda_{qcd}$, but the
Vafa-Witten theorem \cite{Vafa:1984xg} shows that the transition can
only occur in a region where the two flavors have opposite signs for
their masses, i.e. $|m_1|<|m_2|$.

Note that this transition occurs when both $m_u$ and $m_d$ are
non-vanishing but of opposite sign.  At the transition the correlation
length diverges.  This is a simple example of how it is possible to
have significant long distance physics without small Dirac
eigenvalues.  Complimentarily, there is no structure at points where
only one of the quark masses vanishes.  In this situation there is no
long distance physics despite the possible existence of small Dirac
eigenvalues.  This is connected with the difficulty in defining a
vanishing quark mass as discussed in Section \ref{spinflip}.

\begin{figure}
\centering
{\includegraphics[width=4.5in] {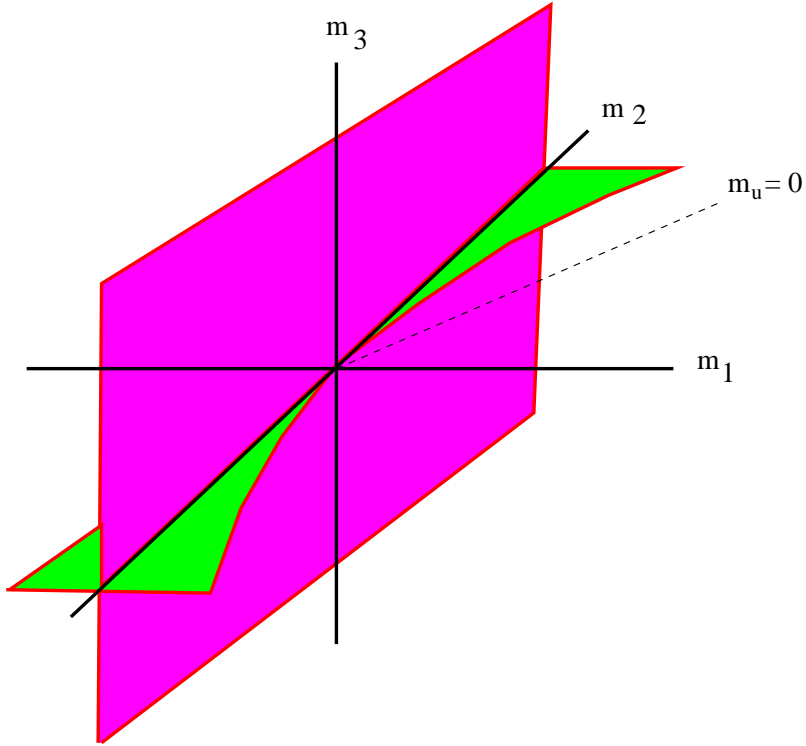}}
\caption{The full phase diagram for two-flavor QCD as a function of
  the three mass parameters.  It consists of two intersecting
  first-order surfaces with second-order edges along curves
  satisfying $m_3=0$, $|m_1|<|m_2|$.  There is no structure along the
  $m_u=0$ line except when both quark masses vanish.}
\label{phasediagram}
\end{figure}

Putting this all together gives the final phase diagram sketched in
Fig.~\ref{phasediagram}.  There are two intersecting first-order
surfaces, one at $\{m_1=0$, $m_3\ne 0\}$ and the second at
$\{m_1<m_2$, $m_3=0\}$.  The latter ends at second-order curves that
touch the lines of vanishing quark mass only at the origin.  The
transition at the origin itself is, of course, that of the four
dimensional $O(4)$ sigma model.  The octets defined by the signs of
the three mass terms are characterized by the signs of the expectation
values for the conjugate fields $\sigma,\pi_0,\eta^\prime$.  The
flavored chiral symmetry of Eq.~(\ref{flavorrot}) combined with
permutation symmetry for the two flavors shows that the eight
corresponding regions divide into two sets of four with equivalent
physics, the sets differing in the sign of CP violating effects.

The first-order surfaces both occur where the formal parameter
$\Theta$ takes the value $\pi$. However, note that with non-degenerate
quarks there is also a finite $\Theta=\pi$ region with $m_2$ near
$m_1$ where there is no transition.  The absence of any physical
singularity at $m_u=0$ when $m_d\ne 0$ lies at the heart of the
problem of defining a vanishing quark mass.

\section{Summary}
\label{summary}
Non-perturbative effects in QCD couple the renormalization group flow
for the masses of different fermion species.  This effect is absent in
perturbation theory, but is automatically included in lattice gauge
simulations.  This coupling means that quark mass ratios are generally
not constants but depend on renormalization scale.  This is true for
vanishing as well as non-vanishing quark masses.  One practical
consequence is that it is inappropriate to match lattice and
perturbative masses.

Taking into account the possibility of CP violation, the general
two-flavor theory depends on 3 mass parameters.  A simple effective
Lagrangian approach reveals an intricate phase diagram containing both
first and second order transitions as the mass parameters are varied.
This diagram displays no structure at $m_u=0$ when $m_d\ne 0$,
suggesting that $m_u=0$ is not an appropriate solution to the strong
CP problem.

\section*{Acknowledgements}
 I am grateful to the Alexander von Humboldt Foundation for supporting
 visits to the University of Mainz where part of this study was
 carried out.  This manuscript has been authored under contract number
 DE-AC02-98CH10886 with the U.S.~Department of Energy.  Accordingly,
 the U.S. Government retains a non-exclusive, royalty-free license to
 publish or reproduce the published form of this contribution, or
 allow others to do so, for U.S.~Government purposes.

\end{document}